\font\caps=cmcsc10 at 12pt
\font\eightrm=cmr8 at 8pt
\documentclass{oxarticle}
\usepackage{amsmath, amssymb}

\newcommand{\cA}{{\cal A}}

\newcommand{\EMI}{Extraordinary Mass Invariant}

\newcommand{\EI}{Extraordinary Invariant}

\newcommand{\PB}{BRST Poisson Bracket}

\newcommand{\dB}{\d_{{}_{{}_{\rm BRST}}}}

\newcommand{\LT}{\LaTeX}

\newcommand{\bt}{\begin{tabular}{c}}
\newcommand{\et}{\end{tabular}}

\newcommand{\np}{\newpage} 
 
\newcommand{\eb}{\ee\be } 
 
\newcommand{\bmat}{\lt ( \begin{array} }
\newcommand{\emat}{  \end{array} \rt )}

\newcommand{\ovW}{{\ov W}}

\newcommand{\cP}{{\cal P}}

\newcommand{\oB}{{\ov B}}

\newcommand{\ovV}{{\ov V}}

\newcommand{\oE}{{\ov E}}

\newcommand{\ED}{\end{document}}

\newcommand{\om}{{\ov m}}

\newcommand{\oC}{{\ov C}}

\renewcommand{\a}{\alpha}	
\renewcommand{\b}{\beta}

\renewcommand{\d}{\delta}

\newcommand{\h}{\eta}

\newcommand{\f}{\phi}
\renewcommand{\c}{\chi}

\newcommand{\w}{\omega}

\renewcommand{\S}{\Sigma}
\newcommand{\U}{\Upsilon}

\newcommand{\la}{\label}
\newcommand{\ci}{\cite}

\newcommand{\ds}{\documentstyle}	
\newcommand{\fr}{\frac}

\newcommand{\pa}{\partial}
\newcommand{\ov}{\overline}
\newcommand{\be}{\begin{equation}}
\newcommand{\ee}{\end{equation}}
\newcommand{\ba}{\begin{array}} 
\newcommand{\ea}{\end{array}}
\newcommand{\bea}{\begin{eqnarray}}
\newcommand{\eea}{\end{eqnarray}}
\newcommand{\ra}{\rightarrow}

\newcommand{\Lra}{\Leftrightarrow}
\newcommand{\lt}{\left}
\newcommand{\rt}{\right}

\newcommand{\ben}{\begin{enumerate}}
\newcommand{\een}{\end{enumerate}}
\newcommand{\bitem}{\begin{itemize}}
\newcommand{\eitem}{\end{itemize}}

\newcounter{apple} 
\newcommand{\articlenumber}{\LT 4455.EIinSOSOfinal}

\proofmodefalse
\usepackage{color} 

\begin{document}

\normalsize

\begin{center}

\vspace*{1in}
{  \huge  
An Extraordinary Mass Invariant and an Obstruction \\in a Massive Superspin One Half Model\\ made with a Chiral Dotted Spinor Superfield\\[1cm] }
%

\vspace*{.1in}
%


\renewcommand{\thefootnote}{\fnsymbol{footnote}}

{\caps John A. Dixon\footnote{jadixg@gmail.com;\\dixon@maths.ox.ac.uk}
\\Mathematical Institute\\ Oxford University \\ Oxford, England}
 


{\bf Abstract}
\end{center}

\large

An earlier paper   introduced an action for a  new kind of irreducible  massive superspin  $\fr{1}{2}$ multiplet, using BRST cohomological techniques including `BRST Recycling'.
A mass term was introduced in the earlier paper.  A second mass term is discussed in this paper.  This new mass invariant is an `Extraordinary Invariant'--it has Zinn sources in it.  The natural treatment for this situation is to `Complete the Action' so that the new action yields zero for the \PB.
In the present case, this Completion meets a BRST Obstruction. Setting the coefficient of this `Completion Obstruction' to zero  restores the massive superspin $\fr{1}{2}$  supermultiplet
 with a new mass made from the two mass terms. Usually an Obstruction appears as an Anomaly at one loop perturbation theory, but this is a different mechanism to produce it.  
\Large

{\bf \theapple}.\;\addtocounter{apple}{1}
 Although supersymmetry 
has undergone intense scrutiny   for over 40 years, there are still profound mysteries and unsolved problems .
The chief of these is that, so far, it does not seem to have any experimental relevance 
\ci{Agashe:2014kda}.
However that may be about to change as results at the LHC continue to be reported
\ci{Allanach:2015xga}.
 But it is also very arguable  that we     really do not know what SUSY predicts
 \ci{Agashe:2014kda, Altarelli:2010uu,  Altarelli:2014xxa,
Baer:2006rs}. 
The spontaneous breaking of
 SUSY gives rise to sum rules that are embarrassing for phenomenology, and a huge cosmological constant which is embarrassing for cosmology
 \ci{Baer:2006rs,
Weinberg:1988cp}. 
  
{\bf \theapple}.\;\addtocounter{apple}{1}
The predictions of SUSY are still in doubt for many reasons, and in particular there is still much to learn about the representation theory of SUSY, even in 3+1 dimensions.  Progress in the representation theory of SUSY is being made 
by the adinkra program and other investigations  of Buchbinder and Gates et al.  \ci{Gatesfundremains,Calkinsthinkdiff,Buchbindersupone,Buchbindersupthree,GatesKoutmassless,Gateslinmassdyn}. 
Massive representations of SUSY are clearly related to some of the puzzles of the superstring (See for example 
\cite{Berkovits:1998ua,Berkovits:1997zd}). New efforts at understanding the cohomology of SUSY BRST  are also under way
\ci{Movshev:2014hha,
Movshev:2013hza,
Movshev:2011pr,
Movshev:2009ba,
Movshev:2003ib,
Lin:2015ixa,
Chang:2014kma,
Chang:2014nwa}.

{\bf \theapple}.\;\addtocounter{apple}{1}
Following along in the path of looking for new representations of SUSY, in \ci{one}, a new supersymmetric action for massive superspin $\fr{1}{2}$ was constructed using `BRST Recycling', rather than superspace.
This action contained the component fields of a chiral dotted spinor superfield, which was expected to have interesting cohomology.  Indeed it does, as we shall show here. 	In addition to the mass term discussed in
 \ci{one}, there is another mass term in that theory, and it will be the subject of this paper.
The new mass term $\cA_{\rm E}$ is
a BRST \EI, which means that it is irrevocably dependent on Zinn sources, and that it satisfies 
\be 
\d 
\cA_{\rm E} 
=0
\la{aeis}
\ee
This kind of object has sometimes been called `finding a consistent extension of a BRST theory' and the papers \cite{Brandt:2002pa,Brandt:2001hs,Barnich:1993vg,Barnich:1993pa,dixonYMseed}
 have 
discussed that concept in the  context of various actions.

{\bf \theapple}.\;\addtocounter{apple}{1}
An unusual feature of the present \EI\ is that an attempt to complete the action, so that the new action yields zero for the \PB, meets a `Completion Obstruction' in the present case.  Following the usual BRST reasoning 
\ci{Becchi:1975nq}, this ghost charge one `Completion Obstruction' could also conceivably arise as an Anomaly, but it clearly does not do so in the present free Action.

{\bf \theapple}.\;\addtocounter{apple}{1}
The new \EI\ $\cA_{\rm E}$  here is written explicitly below in equations (\ref{lineoneofei})
to
(\ref{EMIeq})
in the notation of 
\ci{one}.
In this paper we will  go through the exercise of completing the action so that the completed  action still satisfies the original \PB\ in \ci{one}.  To do this we need to first drop  the gauge and ghost fixing action that was used in  \ci{one}, because we will need to change it after the Completion. Then we put the action plus the \EI\ into the \PB, and observe that the \PB\ is no longer zero.
There are two non zero terms: the variation of a Completion Term and also an Obstruction.  We add the Completion Term, and then also constrain the coefficient of the Obstruction to be zero.
At that point we can add a new, more suitable, form of the gauge and ghost fixing action. Then
we look  at the equations of motion of the new theory, and we see how the Completion term and the Constraint act together to restore the action so that it again describes a  massive superspin $\fr{1}{2}$ supersymmetry multiplet, but with a revised mass.
 Then we consider the origin and significance of the above results.

{\bf \theapple}.\;\addtocounter{apple}{1}
From 
 \ci{one},
 let us take the following action
\be
\cA_{\rm Massless} 
= 
\cA_{\rm Kinetic\;\c} 
+\cA_{\rm Kinetic\;\f} 
+\cA_{\rm Zinn\;\c}  
+\cA_{\rm Zinn\;\f}  
+\cA_{\rm SUSY}  
\la{masslessaction}
\ee
This is the full action from that paper\footnote{$\cA_{\rm SUSY}$ is discussed in footnote 4 of  \ci{one}}, but without the mass term $\cA_{\rm Mass\;\c\;\f} $ and without the ghost and gauge fixing action  $\cA_{\rm GGF} $ of that paper.

This gives rise to 
the following nilpotent BRST operator\footnote{This operator will be written in full detail in \ci{SoSocohom}.}:
\be
\d_{\rm Massless} 
= 
\d_{\rm Kinetic\;\c} 
+\d_{\rm Kinetic\;\f} 
+\d_{\rm Zinn\;\c}  
+\d_{\rm Zinn\;\f}  
+\d_{\rm Field\;\c}  
+\d_{\rm Field\;\f}  
+\d_{\rm Susy}  
\la{masslessdelta}
\ee
where
$\d_{\rm Kinetic\;\c} $ arises from functional derivatives of 
$\cA_{\rm Kinetic \;\c}
$, etc. as described in \ci{one}.  
  It is the usual `square root' of the  \PB\ 
$\cP_{\rm Total}[\cA]$ from \ci{one}, evaluated with $\cA \ra \cA_{\rm Massless} $, where $\cP_{\rm Total}[\cA]$ was defined by equation (6)
 of \ci{one}. 
It is nilpotent because 
\be
\cP_{\rm Total} \lt [\cA_{\rm Massless}\rt ]
=0 \Lra \d^2_{\rm Massless} =0
\la{ptotmassless}
\ee
In \ci{one}, we noted that  the following `Ordinary' mass invariant is in the cohomology space\footnote{The relevant operator in  \ci{one}  was simply what we called $\d_{\rm First}$ in equation (15) of that paper.  Whether we included the 
Zinn variation terms of $\d$ that arise from equations of motion from the two  actions 
$\cA_{\rm Kinetic\;\c} $ and
$\cA_{\rm Kinetic\;\f} $ 
was irrelevant, because $\cA_{\rm O}$ does not contain any Zinns.  But it is important to note that these do not give rise to  $\cA_{\rm O}$ as a boundary.  However for the case of $\cA_{\rm E}$ we need to be more careful, and so we define the new operator $\d_{\rm Massless}$ explicitly in the foregoing.} of $
\d_{\rm Massless}$:

\[
\cA_{\rm O}
= \int d^4 x \;\lt \{
m_1\f_{L \dot \a}
  \c_{R}^{ \dot \a}
+
m_1\ov\f_{R  \a}
\ov\c_{L}^{  \a}
\rt.
\]
\be
\lt.
+m_1 E \ov B
+m_1 W_{\a\dot \a}
\ov V^{\a\dot \a}
+m_1 \eta' 
\ov\w
\rt \}
+ *
\la{omassterm}
\ee

Now we claim that there is another kind of mass term here. The following `Extraordinary' mass invariant is {\bf also} in the cohomology
 space\footnote{Finding this term $
\cA_{\rm E}$ is more tricky than finding the mass term above, as is obvious from its complicated form. 
} 
of $
\d_{\rm Massless}$:

\be
\cA_{\rm E}= 
\int d^4 x 
\lt \{
2 m_2 
\U
\ov \w
- \fr{m_2}{2}
\pa_{\a \dot \a} 
\ovV^{\a \dot \a} 
E
- m_2
Z_L^{ \dot \a}
C^{\a}
\ovV_{ \a \dot \a}
\rt.
\la{lineoneofei}
\ee
\be
+ m_2
\ov Z_R^{ \a}
\oC^{ \dot\a}
\ovV_{ \a \dot \a}
+ m_2
\f_{L \dot \a}
\c_{R}^{ \dot \a}
-m_2\ov\f_{R  \a}
\ov\c_L^{ \a}
\ee
\be
\lt.
- m_2
\S^{\a \dot \a}
\oC_{\dot \b}
\ov\c_{L  \a}
+ m_2
\S^{\a \dot \a}
\c_{R \dot \a}
C_{\a}
+2 m_2
 J' \oB
\rt \} + * 
\la{EMIeq}
\ee

Like the mass term $
\cA_{\rm O}$, the existence of $\cA_{\rm E}$ is indicated by spectral sequence techniques applied to the massless BRST operator $
\d_{\rm Massless}$. 
This somewhat technical analysis will be presented in a third paper \ci{SoSocohom}, where we find even more cohomology than is discussed here\footnote{In fact this theory  contains three independent supersymmetric mass terms and five obstructions. Discussion of the other mass term and the other obstructions would needlessly complicate the present paper.  They do ultimately need analysis of course.}. 

{\bf \theapple}.
\la{theapplehsix}\;\addtocounter{apple}{1}
Note the following:
\ben
\item
The `Ordinary' mass invariant $ \cA_{\rm O}
$ does not contain any Zinn sources.  It contains only fields and Fadeev-Popov ghosts.
\item
The `Extraordinary' mass invariant $ \cA_{\rm E}
$ does contain Zinn sources, namely $\U,Z_L^{ \dot \a},
\ov Z_R^{ \a},
\S^{\a \dot \a}$ and $
J'$. 
\item
Note that all the Zinn sources in $ \cA_{\rm E}
$ are $\f$ type Zinn sources.
There are no  $\c$ type Zinn sources present in $ \cA_{\rm E}
$.
\item
Each term of each invariant contains one $\c$ field.
\item
Each term of each invariant contains one $\f$ field or one $\f$ Zinn source. 
\item
The term $ \cA_{\rm O}
$ contains the fermionic mass terms  $m_1
\lt (
\f_{L \dot \a}
  \c_{R}^{ \dot \a}
+
\ov\f_{R  \a}
\ov\c_{L}^{  \a}
\rt )
$ with a plus sign, 
but the term $ \cA_{\rm E}
$ contains the fermionic mass terms   $m_2
\lt (
\f_{L \dot \a}
  \c_{R}^{ \dot \a}
-\ov\f_{R  \a}
\ov\c_{L}^{  \a}
\rt )
$ with a minus sign.
\la{listofstuff}
\een

{\bf \theapple}.\;\addtocounter{apple}{1}
 If one takes only  
$ \cA_{\rm O}
$ as the mass term, one can just add it to the action and proceed without further ado.  This is because the \PB\ is still zero when one adds  an ordinary invariant.  Since $\cA_{\rm O}$ depends only on Fields, it follows that 

\be 
\cA_{\rm Ordinary}
=
\cA_{\rm Massless}
+\cA_{\rm O}
\ee
satisfies
\be \cP_{\rm Total} \lt [\cA_{\rm Ordinary}\rt ]
=0  
\ee
This happens because of 
(\ref{ptotmassless}) and also because \be
 \cP_{\rm Total} \lt [\cA_{\rm O}\rt ]
=0  
 \ee
 is trivially zero, because $\cA_{\rm O}$ contains no Zinns, and each term of the \PB\ contains one Zinn.
The paper \ci{one}
worked all that out in detail.

{\bf \theapple}.\;\addtocounter{apple}{1}
However things are not so simple when we include the term $ \cA_{\rm E}$ from 
(\ref{EMIeq}) in the action.
This \EMI\ $ \cA_{\rm E}$ gives rise to some new problems, and some new opportunities. So now let us consider the action with both types of mass terms:
\be 
\cA_{\rm ExtraOrdinary}
=
\cA_{\rm Ordinary}
+\cA_{\rm E}
=\cA_{\rm Massless}
+\cA_{\rm O}
+\cA_{\rm E}
\ee

For this case we find, because of the presence of the Zinns in $\cA_{\rm E}$, that the \PB\ is no longer zero.  A simple calculation using the form of the \PB\ from \ci{one} yields:

\be
\cP_{\rm Total} \lt [
\cA_{\rm ExtraOrdinary}
\rt ]
=
2 \cA_{\rm O} \star \cA_{\rm E}
+
\cA_{\rm E} \star \cA_{\rm E}\ee

where

\[
2 \cA_{\rm O} \star \cA_{\rm E}
=
\int d^4 x 
\lt \{
 m_2  
 \om_1   
+
 \om_2  
m_1 \rt \}
\]
\be
\lt \{ 
2\ov \w B
+C^{\a}
V_{ \a \dot \a}
 \c_R^{\dot \a}
-\oC^{\dot\a}
V_{ \a \dot \a}
 \ov\c_L^{  \a}
\rt \}
+*
\la{obstruction}
\ee

and

\be
\cA_{\rm E} \star \cA_{\rm E}=
\lt (  
\om_2 m_2
\rt )
\int d^4 x \lt \{
\oC^{\dot\a}
 \ov\c_L^{  \a}
+  
C^{\a}
 \c_R^{\dot \a}
+ 
\pa_{\a \dot \a} 
\ov \w 
 \rt \}
V^{\a \dot \a} 
+*
\ee

Because of the analog of the Jacobi Identity for $\cP_{\rm Total}$, together with the fact that both mass terms are cocycles of $\d_{\rm Massless}$, both of the above terms are also cocycles of 
$\d_{\rm Massless}$:
\be
\d_{\rm Massless}
\lt (
 \cA_{\rm E} \star \cA_{\rm E}
\rt )
=
\d_{\rm Massless} 
\lt (
 \cA_{\rm O} \star \cA_{\rm E}
\rt )=0
\ee

It turns out that one of these `Poisson Variations' is a coboundary of $\d_{\rm Massless}$ and the other is in the cohomology space of $\d_{\rm Massless}$,

The coboundary is:

\be
\lt (
 \cA_{\rm E} \star \cA_{\rm E}
\rt )
=
-\d_{\rm Massless}
\cA_{\rm Complection}
\ee
where
\be
\cA_{\rm Completion}
= 
-\om_2 m_2
 \int d^4 x 
V_{ \a \dot \a}
\ovV^{ \a \dot \a}
\la{completeionterm}
\ee

However the other term (\ref{obstruction})
is not a coboundary.
It is  in the cohomology
 space\footnote{This will be shown in \ci{SoSocohom} } of $\d_{\rm Massless}$.

The \PB\ of the new action will be zero if we eliminate the two terms above.  We can remove the first by adding the Completion term.  But this is not possible for the second.  The only way to remove  (\ref{obstruction})
 is to set its coefficient to zero:
\be
\lt \{
 m_2  
 \om_1   
+
 \om_2  
m_1 
\rt \}
=0
\la{constraint}
\ee
So to restore the \PB\ to zero, in the presence of both the mass terms, we need to constrain the two mass parameters as in 
(\ref{constraint}), and we also need to add the completion term
(\ref{completeionterm})
.

{\bf \theapple}.\;\addtocounter{apple}{1}
So at this point we have an action of the form

\be
\cA_{\rm Completed}
=\cA_{\rm Massless}
+\cA_{\rm O}
+\cA_{\rm E}
+
\cA_{\rm Completion}
\ee
and it satisfies the equation
\be
 \cP_{\rm Total} \lt [\cA_{\rm Completed}\rt ]
=0  
 \ee
provided that 
(\ref{constraint}) is true.

{\bf \theapple}.\;\addtocounter{apple}{1}
Now we have completed the action so that it yields zero for the \PB.  However $\cA_{\rm Completed}
$ is still gauge invariant.  So now we must add a gauge fixing action.  As usual, we choose this to be a coboundary of the relevant gauge invariant $\d$.
That $\d$ is now the one appropriate to the completed action with the constraint, which arises from the square root of the \PB\ using the action
 $\cA_{\rm Completed}$.

\be
\cA_{\rm New\;GGF}
=
\d_{\rm Completed}
\int d^4 x \; 
 \lt \{
\ov \eta 
 \lt (
\fr{1}{4}
g L+ \fr{1}{2}
\pa_{\a \dot \a} V^{\a \dot \a} 
- \fr{1}{2}g  m_2 E 
\rt )
\rt \}
+* 
\la{newGGFterm}
\ee
In the above we have chosen the gauge fixing term to remove the cross term 
$- \fr{m_2}{2}
\pa_{\a \dot \a} 
\ovV^{\a \dot \a} 
E$ in line (\ref{lineoneofei}) of $\cA_{\rm E}$, 
by using `the 't Hooft trick' \ci{Taylor:1976ru}.
  The part from the variation of $\h$  expands  (choose real $g$), after a shift and integration to
\be
\cA_{\rm Gauge \;Fixing}=
-\fr{1}{ 2 g}  \int d^4 x \;   
\lt\{
   \pa_{\a \dot \a} \ovV^{\a \dot \a}
 \pa_{\b \dot \b} V^{\b \dot \b}
\rt \}
\la{pavpav}
\ee

plus
\be
\cA_{\rm  Cross\;Terms }=
\int d^4 x \;   
  \fr{1}{ 2}\lt\{
   \pa_{\a \dot \a} \ovV^{\a \dot \a}
 m_2 E 
+
 \om_2 \oE 
  \pa_{\a \dot \a} V^{\a \dot \a}
\rt \}
\la{crossterm}
\ee

plus
\be
\cA_{\rm New\;Scalar \;Mass
}=-\fr{g}{2} \int d^4 x \;   
\lt\{
m_2 \om_2
E 
\oE 
\rt \}
\la{newscalarmass}
\ee

{\bf \theapple}.\;\addtocounter{apple}{1}
From the above we have:\footnote{A factor of $\fr{1}{2 }$ was dropped accidentally in the term $- \fr{1}{2 } g 
 \ov \eta 
 C_{\b} \oC_{\dot \b}
\pa^{\b \dot \b}
\h
$ in  \ci{one}.  It has been restored here. }
 \[
\cA_{\rm Ghost}
=
\int d^4 x \; 
\lt \{
\ov \eta 
\Box   \w
+ \eta 
\Box \ov  \w
- \fr{1}{2 } g 
 \ov \eta 
 C_{\b} \oC_{\dot \b}
\pa^{\b \dot \b}
\h
\rt \}
\]
\be
+\int d^4 x \; 
\lt \{
-\fr{1}{2 } \ov \eta 
\pa_{\a \dot \a}\lt(
   \c_L^{\dot \a}
C^{\a}
+
 \ov\c_R^{\a}\oC^{\dot \a}
\rt) 
-\fr{1}{2 }  \eta 
\pa_{\a \dot \a}\lt(
   \c_R^{\dot \a}
C^{\a}
+
 \ov\c_L^{\a}\oC^{\dot \a}
\rt )
\rt \}
\la{ghostaction}
\ee

It is important to remember that the \EMI\ $\cA_{\rm E}$ in (\ref{EMIeq})
contains the Zinn $\U$ and so it changes the transformations of the $E$ field, and this will affect the ghost action.  
In particular, now we have
\be
\d  E = \fr{\d \cA_{\rm Complete}}{\d \ov \U}
=
2 \om_2 \w+
\ov\f_{R \b}
C^{ \b}
-
 \f_{L \dot \b}
\oC^{\dot \b}
\ee
and so we get the following from the term
$
- \fr{1}{2}g  m_2 E 
$ in the action
 (\ref{newGGFterm}).
\be
\cA_{\rm New\;Ghost}
=-
\int d^4 x \;
\ov \h 
\fr{1}{2} \lt[ 
 m_2 \lt ( 
2 \om_2 \w +\ov\f_{R \b}
C^{ \b}
-
 \f_{L \dot \b}
\oC^{\dot \b}
\rt ) 
\rt ]
 +*
\la{newghost}
\ee

{\bf \theapple}.\;\addtocounter{apple}{1}
\la{eqmt}
So now we finally have the completed and gauge fixed action.  It has the form

\be
\cA_{\rm Final}
= 
\cA_{\rm Completed}
+\cA_{\rm New\;GGF}
\eb
= 
\cA_{\rm Kinetic\;\c} 
+\cA_{\rm Kinetic\;\f} 
+\cA_{\rm Zinn\;\c}  
+\cA_{\rm Zinn\;\f}  
+\cA_{\rm Susy}  
\eb
+\cA_{\rm O}
+\cA_{\rm E}
+
\cA_{\rm Completion}
+
\cA_{\rm  Gauge \;Fixing}
\eb
+
\cA_{\rm Cross\;Terms}
+
\cA_{\rm New\;Scalar \;Mass
}
+\cA_{\rm Ghost}
+\cA_{\rm New\;Ghost}
\ee
and it satisfies the equation
\be
 \cP_{\rm Total} \lt [\cA_{\rm Final }\rt ]
=0  
 \ee
provided that we choose 
\be
 m_2  
 \om_1   
+
 \om_2  
m_1 
=0
\ee

{\bf \theapple}.\;\addtocounter{apple}{1}
Now we want to look at the masses and equations of motion of this action.
To see the equations of motion we take the above, set the Zinns to zero and take functional derivatives with respect to the fields.

{\bf \theapple}.\;\addtocounter{apple}{1}
For the scalar equations of motion we have:

\be
\fr{\d \cA_{\rm Fields}}{\d 
B}
=
-2 
\oB
+
\lt (
\om_1 \ov E 
\rt )
=0
\ee

\be
\fr{\d \cA_{\rm Fields}}{\d 
E}
=
-\fr{1}{2}
 \Box \oE 
+m_1 \ov  B
-
\fr{g}{2}m_2 \lt (
\om_2 \ov E 
\rt )
=0
\ee

Putting these together (in the Feynman gauge $g=-1$) yields

\be
\lt (
 \Box - m_1 \om_1 - m_2 \om_2 \rt )
\oE 
=0
\ee

{\bf \theapple}.\;\addtocounter{apple}{1}
Next we look at the vector boson equations of motion, in the Feynman gauge:
\be
\fr{\d \cA_{\rm Fields}}{\d 
\ovV_{ \a\dot \a} }
=
 - \fr{1}{2 }   
\pa^{\a \dot \a} 
\pa^{\b \dot \b} 
 V_{ \b\dot \b} +
\lt ( \Box V^{\a \dot \a}+
\fr{1}{2}\pa^{\a \dot \a} 
\pa \cdot V \rt ) 
\eb
+m_1
  W^{\a\dot \a}
-
 \lt (
 m_2 \om_2 
 \rt )
V^{\a \dot \a} 
=0
\ee

\be
\fr{\d \cA_{\rm Fields}}{\d 
\ovW_{\a\dot \a}}
= 
W^{\a\dot \a}
+
m_1 V^{\a\dot \a}=0
\la{secondvec}
\ee
Putting these together, 
 for this gauge, we get:

\be
 \Box V^{\a \dot \a}
-
 \lt (
 m_1 \om_1 
+
 m_2 \om_2 
\rt )
V^{\a \dot \a} 
=0
\ee

{\bf \theapple}.\;\addtocounter{apple}{1} 
Next we examine the  ghost equations of motion:

\be
\fr{\d \cA_{\rm Fields}}{\d   \eta 
  }
=
\Box   \ov \w
- \fr{1}{2} g 
 C_{\b} \oC_{\dot \b}
\pa^{\b \dot \b}
\ov\h
-\fr{1}{2 } 
\pa_{\a \dot \a}\lt(
   \c_R^{\dot \a}
C^{\a}
+
 \ov\c_L^{\a}\oC^{\dot \a}
\rt) 
\eb
-\fr{1}{2} \lt[ 
 \om_2 \lt ( 
2m_2 \ov\w + \ov\f_{L \b}
C^{ \b}
-
 \f_{R \dot \b}
\oC^{\dot \b}
\rt ) 
\rt ]
 +*
\ee

\be
\fr{\d \cA_{\rm Fields}}{\d 
  \eta'  }
=
\fr{1}{2}
 \lt (
 \ov\f_L^{  \d}
C_{\d}
+ \f_R^{\dot\d}
\oC_{\dot \d}
\rt )
+
m_1 \ov \w
\la{6one}
\ee
To derive a simple equation for the ghost $\w$ we need to add the following fermionic equations: 
\be
C^{\a}
\fr{\d \cA_{\rm Fields}}{\d 
\ov\c_{R}^{ \a}}
=
C^{\a}\pa_{\a\dot\a}
\c_{R}^{\dot \a}
-\fr{1}{2}C^{\a}\pa_{\a \dot \a} \ov\eta
 \oC^{\dot \a}
-C^{\a} (\om_1
+\om_2 )\ov\f_{L   \a}
\ee
and
  \be
 \oC^{\dot \a}
\fr{\d \cA_{\rm Fields}}{\d 
\c_{L}^{\dot \a}}
=
 \oC^{\dot \a}
\pa_{\a\dot\a}
\ov\c_{L}^{ \a}
- \oC^{\dot \a}
\fr{1}{2}\pa_{\a \dot \a} \ov\eta
 C^{ \a}
- \oC^{\dot \a}
(\om_1 
-\om_2) 
\f_{R \dot \a}
\ee
Then we note that the following combination (in the Feynman gauge $g=-1$)  simplifies to yield the ghost equation of motion:
\be
\fr{\d \cA_{\rm Fields}}{\d   \eta 
  }
+
\om_1 \fr{\d \cA_{\rm Fields}}{\d 
  \eta'  }
+
\fr{1}{2}
C^{\a}
\fr{\d \cA_{\rm Fields}}{\d 
\ov\c_{R}^{ \a}}
+
\fr{1}{2}
 \oC^{\dot \a}
\fr{\d \cA_{\rm Fields}}{\d 
\c_{L}^{\dot \a}}
\ee
\be
=
\lt (\Box - m_1\om_1- m_2\om_2\rt)  \ov \w
=0
\ee

{\bf \theapple}.\;\addtocounter{apple}{1} Finally we look at the fermion equations of motion, which are the trickiest case:
\be
\fr{\d \cA_{\rm Fields}}{\d 
\ov\c_{R}^{ \a}}
=
\pa_{\a\dot\a}
\c_{R}^{\dot \a}
-(\om_1
+\om_2 )\ov\f_{L   \a}
-\fr{1}{2}\pa_{\a \dot \a} \ov\eta
 \oC^{\dot \a}
\la{firstone}
\ee

\be
\fr{\d \cA_{\rm Fields}}{\d 
 \f_{L}^{\dot  \a}
}
=  
\pa_{\a\dot\a}
\ov \f_{L}^{\a}
-(m_1
+m_2 ) \c_{R\dot  \a}
-
\fr{1}{2}  
\ov \eta'  
\oC_{\dot  \a}
+\fr{1}{2}  
\ov \h 
 m_2 
\oC_{\dot\a}
=0
\la{secondone}
\ee

  \be
\fr{\d \cA_{\rm Fields}}{\d 
\c_{L}^{\dot \a}}
=
\pa_{\a\dot\a}
\ov\c_{L}^{ \a}
-\fr{1}{2}\pa_{\a \dot \a} \ov\eta
 C^{ \a}
-(\om_1 
-\om_2) 
\f_{R \dot \a}
\la{thirdone}
\ee

\be
\fr{\d \cA_{\rm Fields}}{\d 
\ov \f_{R}^{  \a}
}
=  
\pa_{\a\dot\a}
\f_{R}^{\dot \a}
-\fr{1}{2}  
\ov \eta'  
C_{ \a}
-(
m_1 
-m_2) 
\ov \c_{L  \a}
-\fr{1}{2}  
\ov \h 
 m_2 
C_{\a}
=0
\la{fourthdone}
\ee

Following the reasoning in \ci{one}, 
we want to eliminate the antighost from these equations, using
\be
\fr{\d \cA_{\rm Fields}}{\d 
\w }
= - \Box \ov \h - \om_1 \ov \h'
\la{etaeq}
+ m_2 \om_2 \ov \h
\ee  We will try to write (\ref{firstone}) and 
(\ref{secondone})
 in the form:

\be
\pa_{\a\dot\a}
\c_{R}^{'\dot \a}
-(\om_1
+\om_2 )\ov\f'_{L   \a}
\la{firstone3}
\ee

and

\be
\pa_{\a\dot\a}
\ov \f_{L}^{'\a}
-(m_1
+m_2 ) \c'_{R\dot  \a}
=0
\la{secondone3}
\ee
by setting 

\be
(\om_1
+\om_2 )\ov\f_{L   \a}
=
\lt \{
(\om_1
+\om_2 )\ov\f'_{L   \a}
+ x_1 
\fr{1}{2}\pa_{\a \dot \a} \ov\eta
 \oC^{\dot \a}
\rt \}
\la{fwfijowf}
\ee
and
\be
(m_1
+m_2 ) \c_{R\dot  \a}
=
\lt \{
(m_1
+m_2 ) \c'_{R\dot  \a}
+ x_2
\fr{1}{2}  
\ov \eta'  
\oC_{\dot  \a}
+x_3
\fr{1}{2}  
\ov \h 
 m_2 
\oC_{\dot\a}
\rt \}
\la{sfddsds}\ee
where the unknown variables 
$x_1,x_2,x_3$ are to be determined.  There is no point in including terms like $+ x_4
\fr{1}{2}  
\ov \eta'  
C_{ \a}
$ in
 (\ref{fwfijowf}), or any $C$ terms in (\ref{sfddsds}), because 
 we want to eliminate $\oC$, not $C$, from the equations  (\ref{firstone}) and 
(\ref{secondone}). Substitution  reveals that the solution is:
\be
x_1 = -1 - \fr{\om_2}{\om_1}
\ee

\be
x_2 = 
0
\ee

\be
x_3 = - \fr{m_1 \om_2}{m_2 \om_1}
  - \fr{\om_2}{\om_1}
\la{x3}
\ee
and
\be
x_3 =  1- \fr{\om_2}{\om_1}
\la{x3next}
\ee
Consistency of equations
(\ref{x3}) and (\ref{x3next})
demands that
\be
m_1 \om_2 +m_2 \om_1=0
\ee
which we recognize to be the same constraint 
(\ref{constraint})
that we needed to eliminate the Obstruction.  Note that equations 
(\ref{firstone3}) and 
(\ref{secondone3})
mean that the squared masses of these two fermions are 
$(m_1 + m_2)(\om_1 + \om_2)$. A similar construction can be done for the other two fermion equations 
(\ref{thirdone})
and (\ref{fourthdone})
, except that we arrive at a mass squared there of
$(m_1 - m_2)(\om_1 - \om_2)$.

{\bf \theapple}.\;\addtocounter{apple}{1}
These two sets of fermion masses look different from the masses of the bosons found above.  But   the constraint means that they are in fact the same because
\be
(m_1 + m_2)(\om_1 + \om_2)= (m_1 - m_2)(\om_1 - \om_2)
=
 \lt (
 m_1 \om_1 
+
 m_2 \om_2 
\rt )
\ee
when the constraint  
$m_1 \om_2 +m_2 \om_1=0
$ is true.

{\bf \theapple}.\;\addtocounter{apple}{1}
So there are two independent mass terms in this theory which look like they will give different masses to the two fermions.  However, 
completion of the action actually only leads to a change in the mass without a change in the nature of the supermultiplet. This theory is `playing' with a breaking of supersymmetry, and  the supersymmety is maintained by the constraint 
\be
 m_1 \om_2 + m_2 \om_1 =0.
\la{repeatconstraint}
\ee
So we have recovered the same massive complex superspin $\fr{1}{2}$ multiplet that we started with, except that the mass has changed.  

{\bf \theapple}.\;\addtocounter{apple}{1}
We have discovered that there are two quite different ways to arrive at a ghost charge one BRST Cohomological Obstruction:
\ben
\item
An Obstruction can arise through the Completion of an Action which has an
 \EI, as it does for the present action.
\item
An Obstruction can arise as an Anomaly at one loop perturbation theory  
\cite{Becchi:1975nq}. Many examples of this are known (See for example  
\cite{anomref}). \een
The usual procedure is that one must ensure that an Anomaly which couples to a current that needs to be conserved should have a zero coefficient, or else the theory will be inconsistent.  
In the present paper we have shown a similar result--we must set the coefficient of the Obstruction to zero so that the theory satisfies the \PB, and then we note that we recover a sensible SUSY action.

{\bf \theapple}.\;\addtocounter{apple}{1}
It is natural to ask why this happens in this theory.  In particular why is there a second mass term of the form
 (\ref{EMIeq}), and why is it an \EI?  Why is the BRST cohomology rather rich in this theory?
The answer to that lies in the BRST recycling that is needed to create this multiplet.  The $J'$ Zinn source has zero ghost number, and dimension one, and it plays an important role here. This will be clearer when we use the spectral sequence to derive the cohomology in  
\ci{SoSocohom}, but essentially it comes from the fact that there is a term 
\be
\int d^4 x 
\lt \{ \pa_{\a \dot \a}
J'
\fr{\d}{\d  \S_{\a \dot \a}}
\rt \}
\ee
in the $\dB$ of this theory, which comes from the BRST recycling of the usual gauge variation term:
\be
\int d^4 x 
\lt \{ \pa_{\a \dot \a}
\w
\fr{\d}{\d  V_{\a \dot \a}}
\rt \}
\ee
This leaves underived $J'$ in the theory just as the ghost $\w$ is left in the theory, and both of these generate cohomology.

{\bf \theapple}.\;\addtocounter{apple}{1}
Clearly the fact that there can be two origins for an Obstruction raises an interesting question: Is there a `Doubly Obstructed' theory where both of these mechanisms exist and give rise to the same Obstruction?  If so, could one cancel the coefficients against each other?  There is no point in speculating about this in the absence of an example, but it does seem worthwhile to look for an example. Note that:
\ben \item
The reason that the two fermions in 
$\cA_{\rm E}$ have a different relative sign for mass compared to $\cA_{\rm O}$
is that 
the
field part of 
$\cA_{\rm E}$ breaks SUSY, and that is then corrected by the presence of the Zinn terms, so that $\cA_{\rm E}$ as a whole is a cocyle\footnote{This kind of reasoning was explained in detail in \ci{jumps}} which satisfes (\ref{aeis}).  In the present case the existence of a SUSY charge is thrown into some confusion because of the \EI, which mixes up the equations of motion with the invariance in such a way that the usual derivation of the Noether current  does  not quite succeed 
\footnote
{The  derivation of the SUSY current and charge can be found in any textbook on SUSY, for example \cite{KBbook}. The problem  in the present case is that the assumption that the symmetry arises through field 
variations is not true here--we also need Zinn variations. Put another way, the action $\cA_{\rm Complete}$ is not invariant under SUSY, so there is no supersymmetry charge that governs the spectrum, but the \PB\ is zero, and that suffices for this kind of action. We see in the above analysis that the \PB\ is very strong, because it restores the superspin $\fr{1}{2}$ SUSY multiplet in this case.}.
\item
In the present case there is a serious  problem with the notion of cancelling an Anomaly against a Completion Obstruction, because the same constraint also arises when we  remove the antighost-fermion mixing which is present in this theory, so that 
(\ref{x3}) and 
(\ref{x3next}) will be consistent. Does that kind of constraint always prevent a cancellation between an Anomaly and an Obstruction even if they both exist in the same theory and could be cancelled otherwise? 
\een

{\bf \theapple}.\;\addtocounter{apple}{1}
If such a `Doubly Obstructed' theory exists, it seems likely that it will
be a theory in which there does not exist a set of auxiliary fields that can close the algebra and yield a nice superspace treatment. Certainly whenever one integrates auxiliary fields, in a theory which has them, this will introduce quadratic terms with Zinn sources into the theory.  In such cases, typically one can expect to get a boundary term like the one above in equation 
(\ref{completeionterm}). 
But when will there also be an Obstruction that arises from that completion?  If the theory has a nice superfield treatment, then these boundary terms are artificial in a way, because we can avoid them by keeping the auxiliaries. So it would be a surprise if a `Double Obstruction' appeared in such a theory.

{\bf \theapple}.\;\addtocounter{apple}{1}
There are of course lots of theories where no auxiliary fields exist. This  happens frequently in the more complicated SUSY theories and in higher dimensions.  It also probably happens in the present theory, where we had to use BRST Recycling to obtain the action.
It looks unlikely that the present action possesses auxiliary fields that can generate a nice superfield treatment and restore  the chiral dotted spinor superfield, because it is doubly constrained.  

{\bf \theapple}.\;\addtocounter{apple}{1}
Quite aside from any speculation about `Double Obstructions' we now clearly have a new kind of irreducible supersymmetry multiplet here, and it is stable even when we add the \EMI, provided we impose the necessary constraint, as shown above.  A natural question is whether it can be used in a new kind of SUSY extension for the Standard Model.   If the conserved global phase in  the fields and Zinns $\f,\c\ \cdots$ is taken to be Lepton number, for example, then to recover the three known spin $\fr{1}{2}$  Leptons, we would need to add one pair of chiral scalar superfields.  Can this new superspin $\fr{1}{2}$ action be coupled to supersymmetric gauge theory?  Can it be coupled to Higgs chiral scalar multiplets?  The answers appear to be yes, but without a superspace version, this requires detailed analysis. 
Those  questions are under investigation. Certainly it is very peculiar to have terms like $
\int d^4 x \; \fr{1}{2}
 \eta'  \lt (
 \ov\f_L^{  \d}
C_{\d}
+
 \f_R^{\dot\d}
\oC_{\dot \d}
\rt )
$ in the kinetic action $\cA_{\rm Kinetic \;\f}$ for the $\f$ field, but the results of this paper and of \ci{one} indicate that this is not a problem, and that indeed this term fits nicely into the fermion and ghost actions of the theory, and that it is necessary to keep the BRST invariance.


\vspace{.5in}

\begin{center}
 {\bf Acknowledgments}
\end{center}
\vspace{.2in}

  I thank  Carlo Becchi, Friedemann Brandt, Cliff Burgess, Philip Candelas, Rhys 
Davies, John Moffat, Pierre Ramond, Kelly Stelle, Peter Scharbach,  Raymond Stora,  
J.C. Taylor, Xerxes Tata and Peter West for stimulating correspondence and conversations.


\tiny \articlenumber
\\
\today

\np

\end{document}